\documentclass[reprint,twocolumn,pra]{revtex4-1}
\usepackage{amsmath,amssymb}
\usepackage{epsfig}
\usepackage{xcolor,datetime}
\usepackage[pdfstartview=FitV]{hyperref}

\newcommand{\dg}{\dagger}
\renewcommand{\a}{\ensuremath{\alpha}}
\newcommand{\p}{\psi}
\newcommand{\ket}[1]{\ensuremath{\lvert #1 \rangle}}

\newcommand{\mean}[1]{\ensuremath{\langle #1 \rangle}}
\newcommand{\abs}[1]{\ensuremath{\lvert #1 \rvert}}

\begin{document}
\title{Driven Bose--Hubbard Model with a Parametrically Modulated Harmonic Trap }

\author{N. Mann$^1$, M. Reza Bakhtiari$^1$, F. Massel$^2$, A. Pelster$^3$ and M.
Thorwart$^{1,4}$}
\affiliation{$^1$I.\ Institut f\"ur Theoretische Physik, Universit\"at
Hamburg,
Jungiusstra{\ss}e 9, 20355 Hamburg, Germany\\ 
$^2$Department of Physics and
Nanoscience Center, University of Jyvaskyla,
P.O. Box 35 (YFL), FI-40014 University of Jyvaskyla, Finland \\
$^3$Physics Department and Research Center OPTIMAS, Technical University of
Kaiserslautern, \\
Erwin-Schr\"odinger Stra{\ss}e 46, 67663 Kaiserslautern, Germany \\
$^4$The Hamburg Centre for Ultrafast Imaging, Luruper Chaussee 149, 22761
Hamburg, Germany}


\begin{abstract}
We investigate a one-dimensional Bose--Hubbard model in a parametrically driven global harmonic trap.
The delicate interplay of both the local interaction of the atoms in the lattice  and the driving of the global trap allows us to control the dynamical stability of the trapped quantum many-body state.
The impact of the atomic interaction on the dynamical stability of the driven quantum many-body state is revealed in the regime of weak interaction by analyzing a discretized Gross-Pitaevskii equation within a Gaussian variational ansatz, yielding a Mathieu equation for the condensate width.
The parametric resonance condition is shown to be modified by the atom interaction
strength. In particular, the effective eigenfrequency is reduced for growing interaction in the mean-field regime. 
For stronger interaction, the impact of the global parametric drive is 
determined by the numerically exact time-evolving block decimation scheme.
When the trapped bosons in the lattice are in a Mott insulating state, the absorption of energy from the driving field is suppressed due to the strongly reduced local compressibility of the quantum many-body state.
In particular, we find that the width of the local Mott region shows a breathing 
dynamics.
Finally, we observe that the global modulation also induces an effective time-independent inhomogeneous hopping strength for the atoms.
\end{abstract}


\maketitle
\section{Introduction}
Strong external time-dependent driving is known to have pronounced
implications for quantum many-body systems \cite{Bukov2015B}.
For instance, light can induce a collapse of long-range ordered charge-density-wave phases \cite{schmitt_transient_2008,yusupov_coherent_2010,hellmann_ultrafast_2010,
rohwer_collapse_2011}, deconstruct insulating phases
\cite{rini_control_2007,hilton_enhanced_2007,liu_terahertz-field-induced_2012-1}, 
break Cooper pairs \cite{demsar_pair-breaking_2003,graf_nodal_2011,smallwood_tracking_2012,matsunaga_higgs_2013}, or
induce novel transient superconducting phases
\cite{fausti_light-induced_2011,mankowsky_nonlinear_2014,
hu_optically_2014,kaiser_optically_2014,foerst_2015,singla_2015,
mitrano_optically_2015}. 
An interesting class of externally driven systems are
parametric oscillators in which the characteristic 
frequency is periodically modulated. Already the classical Kapitza
pendulum is known for its peculiar dynamics
\cite{Citro2015} which is stabilized by properly choosing the 
driving parameters. The parametric quantum harmonic
oscillator has even a nonlinear Floquet spectrum
\cite{Zerbe,Thorwart} with regimes of stable and unstable quantum
dynamics. 

Novel concepts of driven quantum many-body systems can be studied in atomic
quantum gases, see Refs.~\cite{Polkovnikov2011,Langen2015,Eisert2015,Torma_book,Eckardt2016} for recent reviews. 
A trapped Bose--Einstein condensate (BEC) with weak interactions is well described by the mean-field Gross--Pitaevskii (GP) equation. In absence of any additional optical lattice, a homogeneous BEC in a time-dependent set-up has been considered in different constellations for a long time. In an early work, Castin and Dum analytically studied a homogeneous BEC in a parametrically modulated  harmonic
trap \cite{add1}. They showed that the driving induces a parametric instability in the global motion of the condensate which gets depleted exponentially fast and 
noncondensed modes become dominantly populated due to this effective ''heating''. The effect of a parametrically  driven trap potential was also studied in Ref.~ \citep{add2} within the GP approach. It was shown that the dynamics of the condensate wave function is described by the classical Mathieu equation of a parametrically forced oscillator, by which one obtains stability criteria. In another 
mean-field study, the effect of a time-dependent scattering length on the collective motion of a BEC was studied in Ref.~\cite{add3}.

In the presence of an optical lattice, the BEC is described by the Bose-Hubbard model which is known to have the two distinct phases of a superfluid or a Mott insulator. Jaksch et al.~\cite{add4} have investigated 
the case of a Bose-Hubbard model with a time-dependent lattice depth which leads to a variation of both the 
on-site interaction and the hopping amplitude. Starting out from the superfluid phase, the atoms are driven to the Mott insulator phase and converted there into molecules. Eventually, 
 the melting of the molecular Mott insulating phase produces a molecular superfluid \cite{add4}.

Furthermore,  a periodically modulated local atomic interaction \cite{Bagnato2010}
can stabilize a Bose-Einstein condensate \cite{Saito2003,BagnatoHulet,Cairncross}. Moreover, 
the superfluid-Mott insulator transition can be controlled 
\cite{Holthaus2008,Zanesini2009,Santos,Zhang} or novel synthetic
quantum matter \cite{Goldman2014} can be realized by Floquet
engineering \cite{Holthaus2016,Eckardt2016}. 
Also, anyonic statistics \cite{Tang} might be accessible \cite{EckardtAnyons}. 
Local modulations can 
coherently control the single-particle tunneling in shaken lattices
\cite{Lignier2007}, magnetic frustration \cite{Struck2011}, and effective magnetic fields 
\cite{Struck2012}. Modulated local onsite Bose-Hubbard interactions can lead to 
correlated tunneling \cite{Meinert2016} and artificial gauge 
potentials, and thus to novel topological phases 
\cite{Goldman2016}. All these works commonly rely on the time-periodic
modulation of {\em local\/} parameters.

An interesting regime which is less explored is realized 
when a strongly interacting gas in the Mott phase is exposed to a 
time-dependent external driving of the global trapping potential. 
When a system is driven parametrically, it exchanges energy with 
the driving field, and in
principle can be heated to infinite temperature 
\cite{Dalessio2013,Rigol_email,Kuwahara2016}. On the other hand, the parametric oscillator
 has regions of dynamical stability as well. So the natural question arises how
does strong atomic interaction affect the stability of a globally parametrically
driven quantum many-body system. Can strong short-range interaction stabilize
 a quantum gas in a parametric trap which would otherwise be
unstable? In turn, can we obtain information on the atomic interaction by 
externally tuning the system to an unstable dynamical state?

In this work,  we show that a {\it global\/} parametric modulation of the 
trapping potential, which does not have to be tuned to local properties,  
can be used to control the stability of the interacting quantum
gas in an optical lattice. In particular, the global dynamics of the 
quantum many-body system in a
parametrically modulated trap can be stabilized or destabilized by tuning the atomic
interaction strength. Conversely, locating the onset of the instability can be used to determine the
atom interaction strength. To illustrate the mechanism, we investigate the
parametrically driven Bose-Hubbard model with repulsive interaction in two regimes. 
First, we consider the regime of weakly interacting  atoms in the lattice in 
the presence of a parametrically modulated global trap.
This can be treated by a mean-field Gross-Pitaevskii ansatz for
the condensate wave function and is supported by
a numerically exact treatment in terms of the time-evolving block-decimation
(TEBD) method. Second, we aim to investigate the interplay of the strongly interacting 
quantum gas in the Mott regime with an additional parametrically modulated trap. To this end, we have calculated 
the time-dependent dynamics in this regime numerically by the TEBD approach. We find that the parametric 
driving leads to a breathing of the width of a local central Mott region which becomes resonant  
at frequencies which are shifted as compared with the noninteracting case. In the 
Mott regime, energy absorption is increasingly suppressed due to the strongly reduced 
compressibility of the Mott region. 

 After introducing the underlying driven Bose-Hubbard model in Sec. \ref{sec:model}, we present our mean-field analysis for parametric resonance based on a discretized GP equation 
in Sec. \ref{sec:GP}. To go beyond the weak-interacting regime, we show our complementary results for strong interactions based on the exact numerical TEBD in 
Sec. \ref{sec:TEBD}. The connection between periodically-driven harmonic trap and site-dependent hopping is clarified in Sec. \ref{sec:sitehopping}.
We summarize our work in Sec. \ref{sec:concl}. 


%
%
\section{Model} \label{sec:model}
We consider a one-dimensional Bose-Hubbard model with
a global harmonic potential with a time-dependent curvature
$V(t)=V_0+\delta{V}\sin\Omega{t}$. The potential has a time-averaged curvature
$V_0$, which is parametrically modulated with the strength $\delta{V}$ and 
 the frequency $\Omega$. The model Hamiltonian reads with $\hbar=1$,
\begin{align}\begin{split}
H(t) =& -J\sum_{\ell}\left(b^{\dg}_{\ell}b_{\ell+1}+\text{H.c.}\right)+\frac{U}{2}\sum_{\ell} n_{\ell} (n_{\ell}-1) \\&+ V(t) \sum_{\ell} ({\ell} - {\ell}_0)^2 n_\ell\,,
\end{split}
\label{eq:modelH}
\end{align}
where $J$ is the hopping amplitude and $U$ the on-site interaction
strength. Furthermore, ${b}_\ell ({b}^\dg_ \ell)$ are the bosonic
annihilation (creation) operators at site $\ell$, and ${n}_\ell= b^\dg_\ell
b_\ell$ denotes the local occupation number operator. We consider a lattice with $M$ sites 
loaded with $N$ bosonic atoms. Thus, the lattice
center is located at $\ell_0=(M-1)/2$. 

\section{Quantum many-body parametric resonance in the mean-field regime} \label{sec:GP}
In the noninteracting limit $U=0$ and for no driving, the
system can be mapped exactly to a discretized quantum harmonic
oscillator with frequency $\omega_0= 2\sqrt{JV_0}$ and unity mass with a Gaussian 
ground state. When the parametric driving 
is switched on, the parametric resonance at
$n\Omega=2\omega_0$ produces regions of instability in the parameter
space \cite{Zerbe,Thorwart} with diverging position and momentum
variances. The driven single-particle problem is still exactly solvable in terms of
the Mathieu equation with its known stability diagram. For 
interacting particles, this is no longer possible. To elucidate the impact of
quantum many-body interactions on the parametric resonance, we first consider 
the mean-field regime.
\subsection{Mean-field description}
A dilute, weakly interacting atom gas at zero temperature is described by the 
mean-field Lagrangian density 
\begin{align}
L =& \frac{1}{N}\sum_\ell\biggl[ \frac{i}{2} (\p_\ell^* \partial_t \p_\ell -\p_\ell \partial_t \p_\ell^* ) + J(\p^*_\ell\p_{\ell+1}+\text{c.c.})\nonumber\\
& - V(t)(\ell-\ell_0)^2 \p_\ell^*\p_\ell-\frac{U}{2}\p_\ell^*\p_\ell^*\p_\ell\p_\ell\biggr],
\end{align}
obtained from the Hamiltonian \eqref{eq:modelH}, 
with the mean-field condensate wave function  $\ket{\psi}=\sum_\ell \p_\ell
b^\dg_\ell \ket{0}$.
By extremizing the Lagrangian with respect to $\p_\ell$($\p_\ell^*$), one 
arrives at a discretized version of the Gross-Pitaevskii equation. To obtain 
an analytic approximation for the
time evolution of the boson gas, we use a Gaussian trial wave function  \cite{Zoller}
\begin{equation}
 \p_\ell = \left(\frac{N^2}{\pi\a^2}\right)^{1/4} \exp\left[{
-\frac{(\ell-\ell_0)^2}{2\alpha^2}}+i\beta(\ell-\ell_0)^2\right]\,,
\end{equation}
with a time-dependent width $\a\equiv\a(t)$ and  
$\beta\equiv\beta(t)$
and minimize $L$ with respect to  $\a$ and $\beta$.

In the following, we consider the regime $J\gg V_0$, where the condensate is 
extended over many sites, i.e., $\a\gg1$. Then, all sums over 
$\ell$ can be approximated by continuous integrals and the Lagrangian takes the form
\begin{equation}
L=2Je^{-\frac{1}{4\a^2}-\beta^2\a^2}-\left[\dot{\beta}+V(t)\right]\frac{\a^2}{2}-\frac{UN}{2\sqrt{2\pi}\a}\,.
\end{equation}
The Euler-Lagrange equations $\partial_x L = 
\frac{d}{dt}\partial_{\dot{x}}L$ for $x=\a , \beta$ provide the equations of motion $\dot{\a} = 4 J_{\gamma}\a\beta$ and
\begin{equation}
\ddot{\a}+\dot{\gamma}\dot{\a}+4J_\gamma{V}(t)\a= \frac{4J_\gamma^2}{\a^3}+\frac{2J_\gamma{U}N}{\sqrt{2\pi}\a^2}\,,
\label{eq:euler2}
\end{equation}
with $\gamma=\frac{1}{4\a^2}+\a^2\beta^2$ and $J_\gamma=Je^{-\gamma}$.

%
%

Aiming at a linear stability analysis, we expand the width $\a(t)=\a_0+\delta\a(t)$ in terms of
small deviations $\delta\a(t)$ from its equilibrium value $\a_0$, i.e., we assume $\a_0\gg\abs{\delta\a(t)}$, 
and linearize Eq.~\eqref{eq:euler2} with respect to $\delta\a$. 
Taking into account correspondingly $\beta(t)=\beta_0+\delta\beta(t)$ with $\beta_0=0$
we find that the stationary solution $\a_0$ follows implicitly from 
$
2V_0\a_0^4=2Je^{-1/4\a_0^2}+{UN}\a_0/{\sqrt{2\pi}}
$
and that the deviation $\delta\a$ from equilibrium  obeys
\begin{equation}
 \delta\ddot{\a}+4J' \left[V' +\delta{V}'\sin\Omega{t}\right]\delta\a =
-4J'\delta{V}\a_0\sin\Omega{t}\,,
 \label{eq:da_math}
\end{equation}
where both the hopping $J'=Je^{-1/4\a_0^2}$ and the driving strength 
$\delta{V}'=\delta{V}(1+1/2\a_0^2)$ are renormalized. Furthermore, 
the atomic interaction 
renormalizes the potential curvature such that 
\begin{figure}
\begin{center}
\includegraphics*[width=\columnwidth]{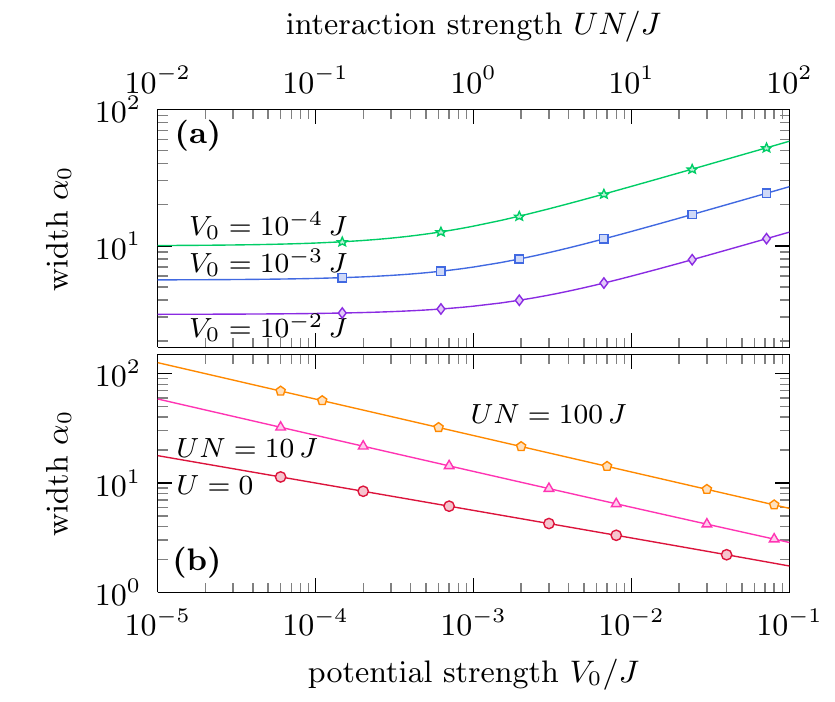} 
\caption{Stationary width $\a_0$ of the static condensate (within the mean-field picture) 
as function of (a) interaction strength $UN$ for different potential curvatures $V_0$, and (b) $V_0$ for different $UN$ ($\delta V=0$).}\label{fig:var}
\end{center}
\end{figure}
\begin{align}\begin{split}
V'=&V_0\left(1+\frac{1}{2\a_0^2}\right)+\frac{J'}{\a_0^4}\left(3-\frac{1}{\a_0^2}
\right)\\&+\frac{UN}{\sqrt{2\pi}\a_0^3}\left(1-\frac{1}{4\a_0^2}\right) \, .\label{eq:res}
\end{split}\end{align}
It  ranges from
$V'\simeq4V_0$ in the non-interacting limit to $V'\simeq3V_0$ in the
Thomas-Fermi limit, i.e., when the kinetic term can be neglected. 
Equation~\eqref{eq:da_math} is the well-known Mathieu equation
with an additional time-dependent force term. Note that the  
 inhomogeneity does not influence the parametric resonance condition \cite{Cairncross}. Thus,
for $\delta{V}'\ll V'$, $\delta\a(t)$ exhibits a resonant behavior when the parametric resonance
condition $n\Omega=2\omega'$ with the resonance frequency $\omega'=2\sqrt{J'V'}$
is fulfilled for $n=1,2,\ldots$.

\subsection{Static trap}
In Fig.~\ref{fig:var} (a), we show the stationary condensate width $\alpha_0$ as
a function of the interaction strength for different potential curvatures. Below
$UN\lesssim{J}$, the condensate width $\a_0\simeq(J/V_0)^{1/4}$ is mainly
determined by the potential curvature $V_0$ and only gradually increases with $UN$. For $UN>J$, the $U$-term  becomes comparable in size to the 
$J$-term which leads to a steeper growth of $\a_0$. In fact, the condensate width behaves as
$\a_0\sim{U}^{1/3}$ in the Thomas-Fermi limit.
Moreover, in Fig.~\ref{fig:var} (b), we show the condensate width 
$\a_0$ as a function of the static potential curvature $V_0$. 
In all cases, we find an algebraic
decrease of $\a_0\sim{V}_0^{-1/\eta}$ with increasing $V_0$, 
where $3\leq\eta\leq4$. In the non-interacting limit we find
$\eta=4$, whereas in the strongly interacting limit we obtain $\eta=3$.
\begin{figure}
\begin{center}
\includegraphics*[width=\columnwidth]{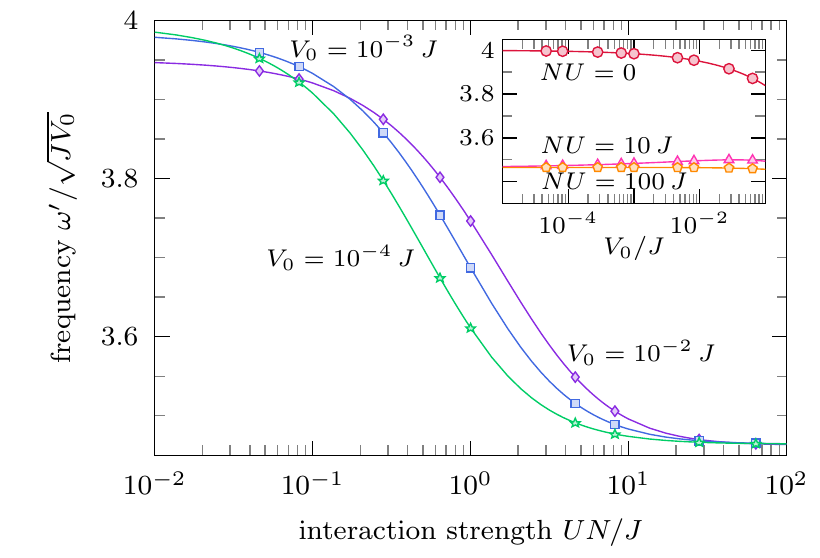} 
\caption{Resonance frequency $\omega'$ as function of $UN$ for different 
$V_0$ (main) and in dependence of $V_0$ for different~$UN$
(inset) for $\delta V=0$.}\label{fig:res}
\end{center}
\end{figure}

The resonance frequency $\omega'$ is also modified by the condensate
interaction. In Fig.~\ref{fig:res} (main), we depict the dependence of 
 $\omega'$ on the interaction strength for various 
potential curvatures. For ease of comparison, we show the resonance
frequency scaled to $\sqrt{JV_0}=\omega_0/2$. For strong interaction $U\gg J/N$, 
the resonance frequency turns out to be independent of the interaction $U$ and
approaches $\omega'=2\sqrt{3JV_0}$, where $J\simeq{J'}$ is satisfied. In the opposite limit of
noninteracting bosons, $J=J'$ is not satisfied {\it per se}. 
This is especially the case for a steep potential when $V_0$ is so 
large that $\a_0$ is only of the order of several lattice sites. In this
regime, the resonance frequency is given by
$\omega'=4\sqrt{JV_0}\sqrt{1+1/(2\a_0^2)} e^{-1/8\a_0^2}$. In the inset of
Fig.~\ref{fig:res}, we show the resonance frequency as a function of the potential
curvature for different interaction strengths.
For a small (large) enough interaction strength $U$ the resonance frequency $\omega'$ weakly decreases
(increases) with the potential steepness $V_0$.

\subsection{Parametrically driven trap}
Next, we address the condensate stability in the parametrically modulated trap. 
To get information about the onset of the parametric instability
on a general basis, we write Eq.~\eqref{eq:da_math} in terms of two
coupled first-order linear differential equations. By using the Floquet
theorem~\cite{Forghani2016}, we determine whether the solutions for a given set
of parameters are stable or not (see appendix for more details).
In Fig.~\ref{fig:stab}, we show the resulting stability diagram as a function of the
interaction strength $UN$ and the driving strength $\delta{V}$ for a fixed
potential curvature $V_0$ and three different values of the driving frequency
$\Omega$. Each curve divides the parameter space into regions with stable or
unstable behavior of the condensate. In the region below each
curve, all solutions of Eq.~\eqref{eq:da_math} are stable, while in the region above, at least 
one solution is unstable. At 
 resonance, i.e., $n\Omega=2\omega'=4\sqrt{J'V'}$,
  an infinitesimally small driving amplitude is sufficient to
destabilize the condensate entirely. A finite particle
interaction may cause a transition from a stable to an unstable behavior. 
Hence, atom-atom interaction may be explored to stabilize a condensate in
a parametrically modulated trap by modifying the resonance condition. Moreover,
the onset of the parametric instability can be used to measure
 the interaction strength.

\begin{figure}
\begin{center}
\includegraphics*[width=\columnwidth]{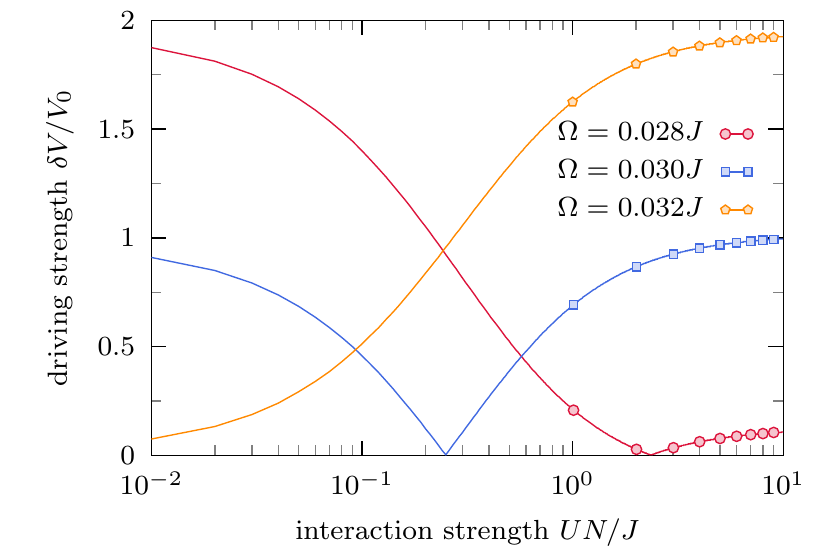} 
\caption{Stability diagram of parametrically driven BEC for the first resonance $n=1$. 
Horizontal axis indicates the particle interaction $UN$ and  vertical axis the
parametric driving strength $\delta{V}$ for $V_0=(1.6\times10^{-5})\,J$ for
different driving frequencies $\Omega$, as indicated. Unstable (stable) solutions
exist in the regions above (below) each curve.}\label{fig:stab}
\end{center}
\end{figure}

\section{Strongly interacting atoms} \label{sec:TEBD}
Next, we determine the quantum many-body dynamics of a strongly interacting 
gas in a lattice and in a parametrically
driven trapping potential. To this end, we use the numerically exact 
time-dependent TEBD method, which is a variant of the time-dependent 
density matrix renormalization group 
\cite{Vidal2003,Vidal2004,Daley2004}. We determine the numerically
exact transient dynamics and are able to investigate the onset 
of Mott physics and its interplay with the parametric 
resonance.
\begin{figure}[t]
\begin{center}
\includegraphics*[width=\columnwidth]{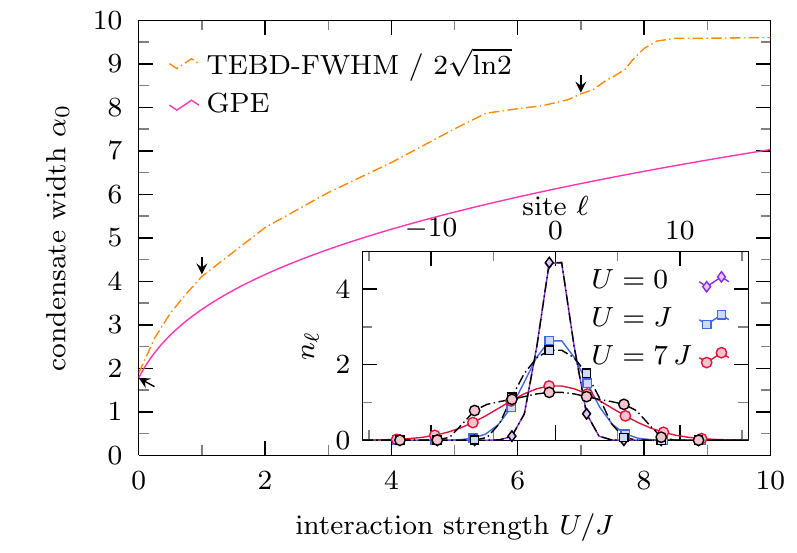} 
\caption{Comparison of the condensate width determined by the  
numerically exact FWHM (TEBD, dashed line) and variational mean-field result $\alpha_0$
(GPE, solid line) for a half-filled lattice of $M=32$ sites. The curvature of
the trapping potential is set to $V_0=0.0922\,J$ and $\delta{V}=0$. The step refers to the 
Berezinsky--Kosterlitz--Thouless quantum phase transition (see text). Inset: Local
occupation number $\langle n_\ell\rangle$ $(\p_\ell^*\p_\ell)$ for 
$U=0$ (diamond), $U=J$ (square) and $U=7\,J$~(circle) for the same $V_0$
(dashed lines are for TEBD, solid lines are for GPE). }\label{fig:comp}
\end{center}
\end{figure}
\subsection{Static trap}
First, we consider the static case $\delta{V}=0$ and calculate the
ground state of the Hamiltonian Eq.~\eqref{eq:modelH}. Then, the condensate
width is extracted as the full width at half maximum (FWHM) of the 
distribution of the local occupation number $\mean{n_\ell}$ \citep{note1}. 
We use a lattice with $M=32$ sites filled with $N=16$ bosons 
in a trap with $V_0=0.0922\,J$. 
In Fig.~\ref{fig:comp}, we show the condensate width in dependence of 
$U$ calculated numerically exactly 
in comparison with the mean-field result $\alpha_0$. The inset depicts  
the corresponding distribution  $\mean{n_\ell}$.  A perfect
agreement is found for $U=0$, where both approaches yield coinciding 
Gaussians. Deviations between the two results increase for growing 
$U$ as expected. For $U=7\,J$, a 
plateau-like Mott region with a nearly integer occupation number 
starts to appear in
the numerical result, see inset of Fig.~\ref{fig:comp}. 
This wedding-cake-like structure cannot be reproduced
by the variational mean-field approach. 
The reasons for the deviations are twofold. Beyond finite-size effects in the numerically exact result, for increasing $U$, the condensate starts to locally form a Mott insulating state. The quantum fluctuations at larger $U$ become more important and lead to a broadening of the population density. This is not taken into account by the GPE mean-field approach, but is captured by the TEBD. In fact, for all $U$, the condensate width is systematically larger by TEBD than predicted by the mean-field approach. Interesting is the step-like increase of the condensate FWHM at $U\approx 7J$ which accompanies the formation of the Mott plateau. In fact, this is a signature of the Berezinsky--Kosterlitz--Thouless quantum phase transition which has been shown to occur between $U=7.5 \,J$ and $U=8\, J$ for the $n=1$ Mott lobe \cite{DMRGexact,Pelster2009}. 
\begin{figure}
\begin{center}
\includegraphics*[width=\columnwidth]{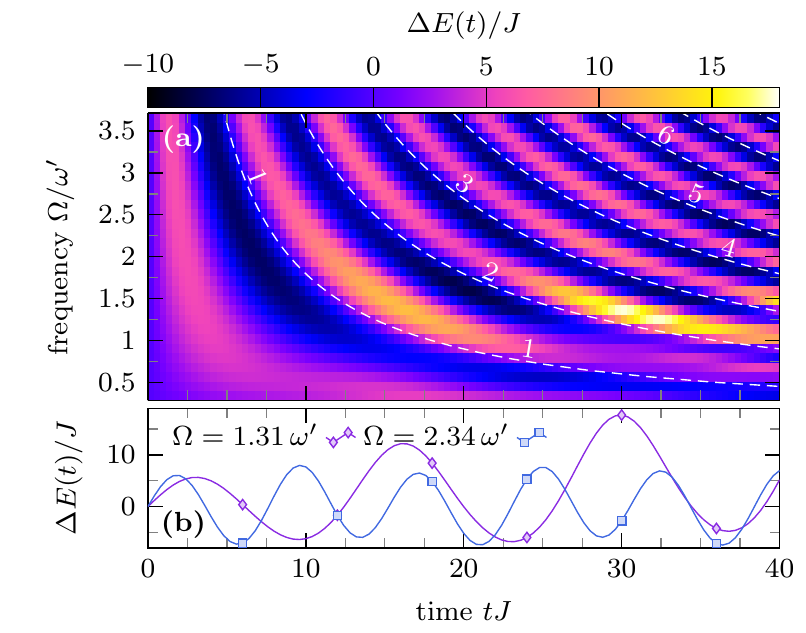} 
\caption{(a) Energy difference $\Delta{E}(t)=E(t)-E(0)$ between  initial
time and $t$ as a function of time $t$ and driving frequency $\Omega$. White dashed lines
mark those times for which $t=2\pi{n}/\Omega$. (b) Cut along constant
$\Omega$-lines as indicated. Parameters are $M=64$, $N=32$, $V_0=0.01\,J$,
$\delta{V}=0.002\,J$, and $U=8\,J$.}\label{fig:tebd}
\end{center}
\end{figure}
\subsection{Parametrically driven trap}
Having studied the static trap, we
next address the transient dynamics of the parametrically driven trap.
At initial time $t=0$, we prepare the system in the ground state of the 
Hamiltonian $H(0)$.   
In Fig.~\ref{fig:tebd} (a), we show the time evolution of the total energy 
$\Delta{E}(t)=E(t)-E(0)$ as a function of time $t$ and driving frequency $\Omega$ with $E(t)=\langle
H(t) \rangle$. In Fig.~\ref{fig:tebd} (b), we show two cuts along the lines
$\Omega=1.31 \, \omega'$ and $\Omega=2.34 \,\omega'$. The resonance frequency
$\omega'$ can be estimated according to Eq.\ \eqref{eq:res}. Resonant driving
in Fig.~\ref{fig:tebd} is expected to occur close to $\Omega/\omega'=2/n$. We
find clear evidence of the two-photon resonance $n=2$ which is manifest in a
growth of the total energy after each period. The resonance is slightly shifted
to $\Omega\simeq1.3 \, \omega'$, due to the strong interaction between the atoms
which is not taken into account in the variational mean-field description. However, the
$n=1$ resonance is not observed in the transient dynamics, but is expected to
occur at longer times. For off-resonant driving, the energy oscillates with the 
frequency $\Omega$ and with small variations in the amplitudes. 

\begin{figure}
\begin{center}
\includegraphics*[width=\columnwidth]{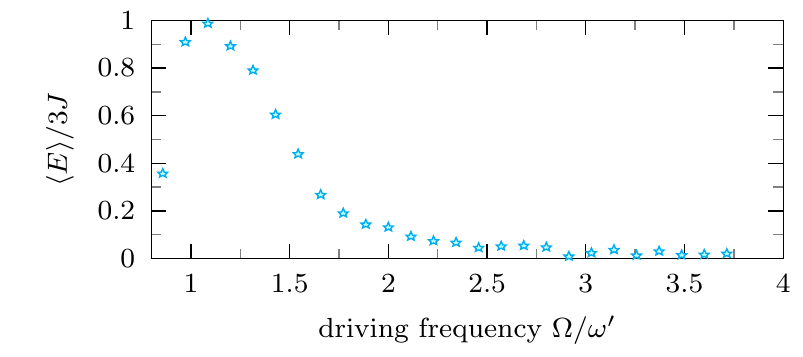} 
\caption{Time-averaged energy $\mean{E}$ as a function of the driving frequency $\Omega$ for the case shown in Fig.\ \ref{fig:tebd}. The same parameters as in Fig.\ \ref{fig:tebd} were used.}\label{fig:mbres}
\end{center}
\end{figure}

For a better quantification of the resonance in the transient dynamics, we define the energy absorption 
over a certain number $m$ of periods according to 
\begin{equation}
  \mean{E} = \frac{\Omega}{2\pi m} \int_{0}^{2\pi{}m/\Omega} dt\; \Delta{}E(t)\,.
  \label{meanen}
\end{equation}
For the set of parameters of Fig.~\ref{fig:tebd}, we are able to calculate the dynamics 
until $t=40/J$. This time window encompasses up to $m=7$ periods over which we take the average in Eq.\ (\ref{meanen}). The result is shown in Fig.~\ref{fig:mbres}.
Note that for smaller values of $\Omega$, the time needed to complete the $m$th period is longer than for larger driving frequencies.
Therefore, data points for smaller $\Omega$ are averaged over fewer periods. 
Yet, this does not change the result substantially. 
A rather broad peak shows up around the second resonance $\Omega\simeq\omega'$ which is slightly shifted to higher frequencies. The slight oscillations at the flank as well as the rise of the first resonance around $\Omega\simeq2\omega'$ can be expected to vanish when longer times are taken into account for the averaging.  
\begin{figure}
\begin{center}
\includegraphics*[width=\columnwidth]{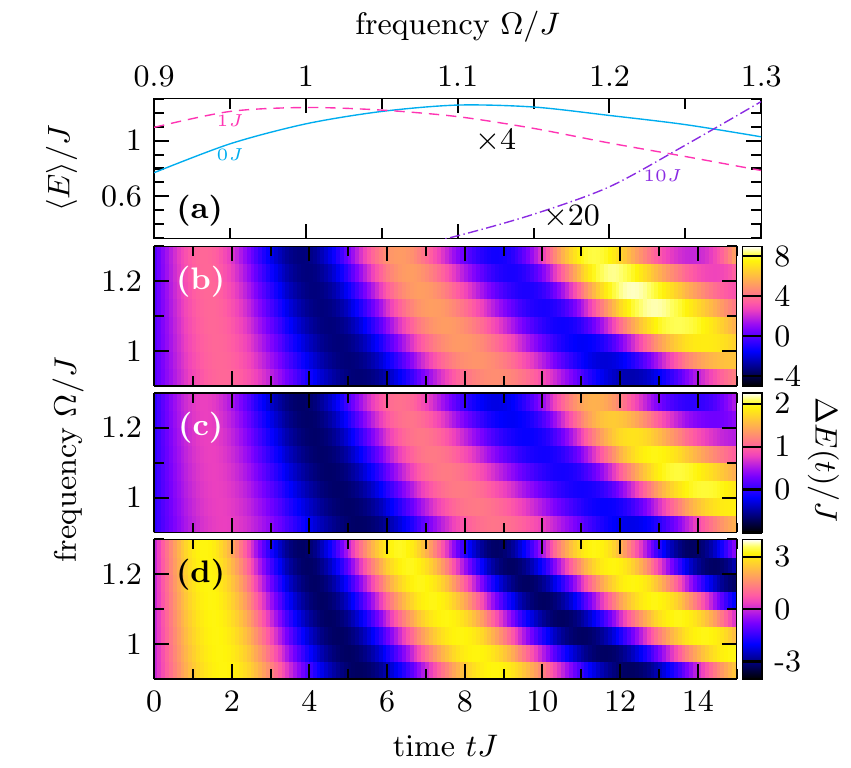} 
\caption{(a) Time-averaged energy $\mean{E}$, calculated over $m=2$ periods, 
 of the energy difference $\Delta{}E(t)$, as shown in panels (b) to (d), as a function of the frequency $\Omega$ for different values of the interaction strength. The solid line corresponds to (b) with $U=0$, the dashed line  to (c) with $U=J$, and the dash-dotted line to (d) with $U=10\,J$. Note that the dashed (dash-dotted) line has been multiplied by a factor of $4$ ($20$) for better illustration. We set $M=32$, $N=16$, $V_0=0.092\,J$ and $\delta{}V=0.01\,J$. 
  The remaining parameters are chosen the same as in Fig.~\ref{fig:comp}.}\label{fig:int}
\end{center}
\end{figure}
%

As can be seen in the inset of Fig.\ \ref{fig:comp}, when the interaction is below or still in the vicinity of the Berezinsky-Kosterlitz-Thouless phase transition, the mean-field result for particle density agrees well with that of TEBD.  However, for stronger interaction, this is in general no longer true and the features of the phase transition are not caught by a mean-field ansatz. In a global harmonic potential, zones of different quantum phases may coexist and sites that locally realize a Mott insulator state hinder the expansion and contraction of the condensate. This can be seen in Fig.~\ref{fig:int} where we show a comparison of the time evolution of the total time-dependent energy $\Delta{}E(t)$ for the cases of non-interacting bosons in (b), interacting bosons with $U=J$ (c), and strongly interacting bosons with $U=10\,J$ (d). While the ground state (with the static potential curvature $V_0=0.0922\,J$) of the cases (b) and (c) occupies a superfluid state on all sites, the central about $14$ sites of the ground state in case (d) occupy a Mott state. According to the mean-field 
result of Eq.~\eqref{eq:res}, the resonance frequencies can be estimated for the three cases to be $\omega'\simeq1.18\,J$ (b), $\omega'\simeq1.06\,J$ (c), and $\omega'\simeq1.05\,J$ (d). We find that the 
 resonance frequencies for the cases (b) and (c), determined by the mean-field argument, fit well to the transient behavior that we observe. This is also reflected in the time-averaged energy shown in Fig.~\ref{fig:int} (a). The slight discrepancies can be traced back to the choice of a rather steep potential such that $\alpha_0\simeq 1$. Then, the mean-field ansatz becomes not very reliable and further corrections become necessary. However,  for 
 the case of strong interaction shown in Fig.~\ref{fig:int} (d), no significant energy absorption occurs  in the vicinity of the resonance frequency and the energy returns to almost its initial value after each period. In contrast, in the cases (b) and (c) a clear growth can be seen.

\begin{figure}
\begin{center}
\includegraphics*[width=\columnwidth]{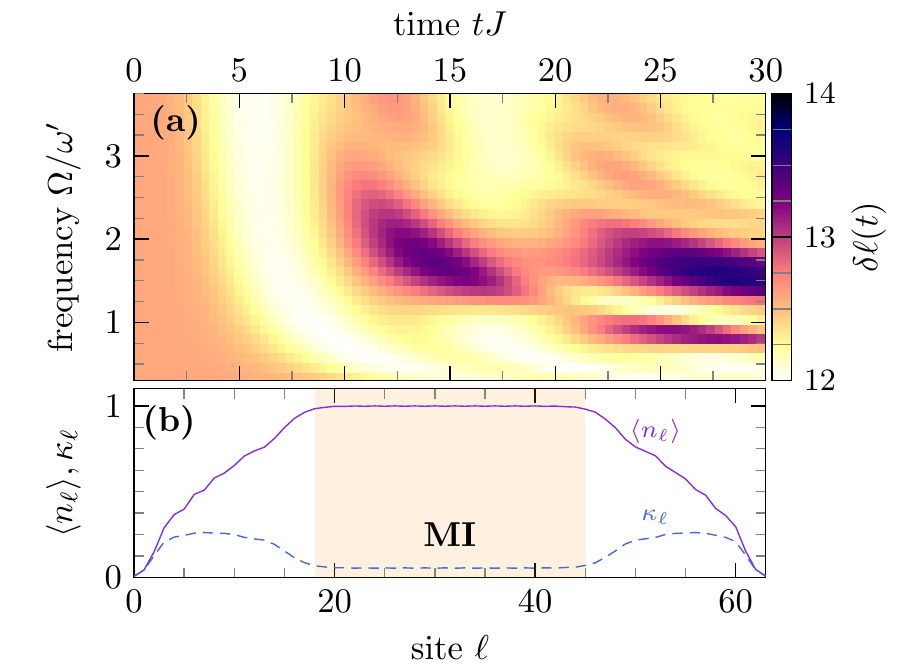} 
\caption{(a) Time evolution of the width $\delta\ell$ of the particle density under a parametric drive. The parameters of the static potential curvature and the driving strength are $V_0=0.01\,J$ and $\delta{}V=0.005\,J$, respectively. We have chosen the time-dependent potential curvature $V(t)=V_0-\delta{}V\cos\Omega{}t$. The other parameters are  $M=64$, $N=48$, and $U=12\,J$. The resonance frequency $\omega'$ is estimated to be $0.346\,J$. (b) Ground-state profile $\langle{n_\ell}\rangle$ of the particle density and local compressibility $\kappa_\ell=\langle{n_\ell^2}\rangle-\langle{n_\ell}\rangle^2$ of the initial Hamiltonian with $V(0)=V_0-\delta{}V$. The shaded region indicates sites within the Mott insulator (MI) phase.}\label{fig:8}
\end{center}
\end{figure}

Further, we show in Fig.~\ref{fig:8} (a) the time evolution of the time-dependent width $\delta\ell(t)=\sum_\ell \mean{n_\ell}_t\lvert \ell-\ell_0\rvert$ of the particle density for different values $\Omega$ of the driving frequency. We choose a parametric drive of the form $V(t)=V_0-\delta{}V\cos(\Omega t)$, such that $V(0)\leq V(t)$ is satisfied for all times, and a rather large driving strength $\delta{}V=0.5\,V_0$ in order to enhance the impact of the parametric resonance. For each frequency, we choose the initial state to be the ground state of the initial Hamiltonian $H(0)$.
The transient behavior shows a growth of $\delta{}\ell$ in the vicinity of $\Omega\simeq2\omega'$ and  $\Omega\simeq\omega'$. To illustrate that the initial state locally occupies a Mott state, we show 
in Fig.~\ref{fig:8} (b) the density profile of the initial state $\mean{n_\ell}$ and the local compressibility $\kappa_\ell=\mean{n_\ell^2}-\mean{n_\ell}^2$. In the central about $25$ sites, the system initially occupies a Mott insulating state, which is characterized by a reduced compressibility. The particle motion on these sites is strongly suppressed because an energy of the order of $U$ is needed to move particles inside this region. In that sense, the  Mott region acts as a local barrier that suppresses the contraction process during the time evolution.
Consequently, particle currents are only observed inside the superfluid regions during the time evolution. 


\section{Effective site-dependent hopping} \label{sec:sitehopping}
The parametric driving of the global 
trap can also be used to create a spatially varying hopping strength.
By a time-dependent unitary transformation $U(t)=\exp[{i\delta{}V\sin(\Omega{}t)\sum_\ell(\ell-\ell_0)^2 n_\ell/\Omega}]$,
the time dependence of the potential can be converted to a time- and site-dependent hopping amplitude. With this, the bosonic annihilation operator transforms according to 
\begin{equation}
  U(t)\,b_\ell\, U^\dagger(t)=b_\ell \, e^{-i\delta{}V\sin(\Omega{}t)(\ell-\ell_0)^2/\Omega}\,.
\end{equation}
The exponential can be absorbed into the hopping amplitude to define a site- and time-dependent hopping amplitude  
\begin{equation}
J_\ell(t) =J\, e^{i\delta{}V\sin(\Omega{}t)[2(\ell-\ell_0)+1]/\Omega}\,.
\end{equation}
Making use of the Jacobi-Anger
identity to expand the exponential in terms of the $m$-th Bessel function of the 
first kind $\mathcal{J}_m(x)$, we obtain  
\begin{equation}
J_\ell(t) = J \sum_m \mathcal{J}_m(\delta V [2(\ell-\ell_0)+1]/\Omega) \, e^{-im\Omega{t}}\,.
\end{equation}
Thus, for a large enough driving frequency $\Omega$, the time average yields an effective local hopping
$J_{\textup{eff}\ell}= J \mathcal{J}_0 (\delta V [2(\ell-\ell_0)+1]/\Omega)$, where the spatial dependence is imprinted by the Bessel function
$\mathcal{J}_0(x)$. 

\section{Conclusions}\label{sec:concl}
 We have studied a driven one-dimensional Bose-Hubbard model with a
periodically modulated harmonic trap. It describes an interacting gas of bosonic atoms 
in a lattice which is placed in a parametrically modulated trapping potential. This 
model allows for the investigation of the interplay of strong atomic interactions in the 
Mott state of the lattice and the external parametric drive. We have 
 analyzed the parametric resonance condition first in the mean-field regime of weak atomic 
 interactions. We find that also in the presence of the lattice, the condensate width 
 is governed by the Mathieu equation. The resonance frequency of the condensate is shifted 
 to lower values for increasing interaction. Moreover, the phase diagram of stable and unstable 
 dynamics is inherited from the Mathieu equation. Furthermore, for stronger interaction, 
 the mean-field approach becomes invalid and the numerically exact TEBD technique has to 
 be invoked. For strong interaction, we observe the formation of a local Mott-insulator 
 region in which the movement of atoms also in the presence of the driving becomes suppressed. 
 Finally, we have demonstrated that the global parametric modulation yields site-dependent 
 hopping amplitudes which can be controlled. Interestingly, locating the onset of the 
 instability allows, in principle, to determine the atom interaction strength. Thus, dynamically probing
a quantum many-body system with a periodic modulation of the harmonic confinement provides a diagnostic tool, which warrants an experimental realization in the realm of ultracold Bose gases.

\begin{acknowledgments}
 This work was supported by the German Research Foundation (DFG), the DFG Collaborative Research Centers SFB 925 and SFB/TR185, and by the Academy of Finland (contract No. 275245). 
 \end{acknowledgments}

\appendix*
\section{Stability analysis}
In this Appendix, we provide more details on the 
stability analysis of Eq.\ \eqref{eq:euler2}. To study the appearance of the instability in the parametrically modulated trap potential, we make use of the well-known Floquet theorem. The theorem is directly applicable to systems whose dynamics are described by a set of $d$ coupled linear differential equations
\begin{equation}
\dot{\bf{x}}=A(t)\bf{x},\label{eq:1}
\end{equation}
where $A(t+T)=A(t)$ is a $d$-dimensional matrix with periodicity $T$.
The main idea is to evolve the fundamental solution ${\Phi(t)}$ over a single period $T$ in time. Whether the system is stable (i.e., all solutions of \eqref{eq:1} are stable) or unstable (i.e., there exists at least one solution of \eqref{eq:1} that is unstable) can be characterized by the eigenvalues $\lambda_{i=1, \dots, d}$ of the monodromy matrix $B=\Phi^{-1}(0)\Phi(T)$. The eigenvalues $\lambda_i$ are directly connected to the Floquet exponents $\nu_i$ via $\lambda_i=e^{T\nu_i}$. 
When all eigenvalues satisfy $\abs{\lambda_i}\leq 1$, the real part of each Floquet exponent is less than or equal to zero and the solution is stable. In turn, in the case that $\abs{\lambda_{i_0}}>1$ for any $i_0$, there exists an unstable solution of \eqref{eq:1} with $\text{Re}(\nu_{i_0})>0$.

To apply the Floquet theorem to Eq.\ \eqref{eq:euler2}, we define the variable $\tau = \sin\Omega{t}$, which allows us to rewrite Eq.\ \eqref{eq:euler2} in terms of a linear differential equation like \eqref{eq:1}
with the vector ${\bf{x}}=(\delta\a,\dot{\delta\a},\tau,\dot{\tau})^t$ and the matrix
\begin{equation}
A(t)=\!\begin{pmatrix}
 0 & 1 & 0 & 0\\
 \!\!-4J'(V'\!+\delta{V'}\sin\Omega{t}) & 0 & -4J'\delta{V}\a_0 & 0\\
 0 & 0 & 0 & 1\\
 0 & 0 & -\Omega^2 & 0
\end{pmatrix}\!,
\end{equation}
with the period $T=2\pi/\Omega$.
To construct the fundamental solution, we have to calculate the time evolution $\textbf{x}^{j=1,\ldots,4}(t)$ of four linearly independent initial vectors ${\bf x}^{j}(0)$. Then, the fundamental solution is given by $\Phi(t)=\begin{pmatrix}{\bf x}^{1}(t) & {\bf x}^{2} (t) & {\bf x}^{3} (t) & {\bf x}^{4} (t)\end{pmatrix}$.

In practice, we choose the initial conditions such that $\Phi(0)={1}$ is the identity matrix. From this, we numerically calculate $\Phi(t)$ and analyze the spectrum of the monodromy matrix $B$. In the eigenvalue spectrum of $B$, we always find two eigenvalue $\lambda_{j_1}=\lambda_{j_2}=1$ which corresponds to the solutions of the differential equation $\ddot{\tau}=-\Omega^2 \tau$. The remaining two eigenvalues are used to characterize the stability behavior of Eq.\ \eqref{eq:euler2}.



\begin{thebibliography}{}
\bibitem{note1}Note that the distribution of the site occupation number $\mean{n_\ell}$ is no 
longer Gaussian.
\end{thebibliography}
\end{document}